\documentclass[acmsmall]{acmart}
\setcopyright{none}
\acmDOI{}

\usepackage{amsmath}
\usepackage{amsfonts}
\usepackage{graphicx}
\usepackage{subfigure}
\usepackage{algorithm}
\usepackage{algorithmic}
\usepackage{multirow}

\begin{document}
	
\title{Understanding the Error in Evaluating Adversarial Robustness}

\author{Pengfei Xia}
\email{xpengfei@mail.ustc.edu.cn}

\author{Ziqiang Li}
\email{iceli@mail.ustc.edu.cn}

\author{Hongjing Niu}
\email{sasori@mail.ustc.edu.cn}

\author{Bin Li}
\authornote{Corresponding author.}
\email{binli@ustc.edu.cn}

\affiliation{
	\institution{University of Science and Technology of China}
	\city{Hefei}
	\state{China}
}

\begin{abstract}
Deep neural networks are easily misled by adversarial examples. Although lots of defense methods are proposed, many of them are demonstrated to lose effectiveness when against properly performed adaptive attacks. How to evaluate the adversarial robustness effectively is important for the realistic deployment of deep models, but yet still unclear. To provide a reasonable solution, one of the primary things is to understand the error (or gap) between the true adversarial robustness and the evaluated one, what is it and why it exists. Several works are done in this paper to make it clear. Firstly, we introduce an interesting phenomenon named gradient traps, which lead to incompetent adversaries and are demonstrated to be a manifestation of evaluation error. Then, we analyze the error and identify that there are three components. Each of them is caused by a specific compromise. Moreover, based on the above analysis, we present our evaluation suggestions. Experiments on adversarial training and its variations indicate that: (1) the error does exist empirically, and (2) these defenses are still vulnerable. We hope these analyses and results will help the community to develop more powerful defenses.
\end{abstract}

\maketitle

\section{Introduction}
Deep neural networks have achieved tremendous success in almost all visual and language tasks \cite{krizhevsky2012imagenet, girshick2015fast, devlin2018bert}. While deep models have also been demonstrated to be vulnerable: models can be easily fooled by perturbing underlying examples with small, imperceptible changes \cite{szegedy2013intriguing, goodfellow2014explaining}. Such crafted inputs, also known as \textit{adversarial examples}, have raised a great challenge to researchers to build safe and robust models for security-sensitive applications, such as biometric identification and self-driving cars \cite{sharif2016accessorize, heaven2019deep}.

Since Szegedy et al. \cite{szegedy2013intriguing} first noticed the existence of adversarial examples in deep models, many efforts have been done to improve the resistance of models against such attacks, e.g., adversarial training \cite{goodfellow2014explaining, madry2017towards, tramer2017ensemble, wang2019bilateral}, gradient regularization \cite{ross2018improving, jakubovitz2018improving} and data preprocessing \cite{gu2014towards, xie2017mitigating, kou2019enhancing}. However, constructing an adversarially robust model is not as easy as previously expected. An appropriate defense should not only deal with various types of attacks but also avoid utilizing ``off-site'' factors, such as gradient masking or obfuscated gradients \cite{papernot2017practical, athalye2018obfuscated}, to build a false sense of security. 

Effectively and reasonably evaluating the adversarial robustness of deep models plays an important role in building an adversarially robust model. An efficacious evaluation procedure helps researchers or practitioners to rationally understand different defense methods, rather than being unnecessary optimistic about incomprehensive assessments. Unfortunately, there is still a lack of unified principles to guide the evaluation. Currently, most of the previous works evaluated their defenses by testing on few attack methods, usually FGSM \cite{goodfellow2014explaining}, PGD \cite{madry2017towards} or C\&W attacks \cite{carlini2017towards}. However, using these attack methods to evaluate is not always sufficient. Needless to say various types of attack, even for a single one, demonstrating that the defense is really effective rather than using gradient masking \cite{papernot2017practical, athalye2018obfuscated} or other ``off-site'' factors is necessary. Some works \cite{athalye2018obfuscated, tramer2020adaptive} also support this view: the authors demonstrated that recently published defenses are still vulnerable against properly performed adaptive attacks.

In order to design an effective evaluation, one needs to understand: (1) what should the adversarial robustness be defined, and (2) what is the error (or gap) between the evaluated adversarial robustness and the expected true one and why the gap exists. After that, an evaluation can be designed according to the error to approximate the true value. Regarding the first issue, Madry et al. \cite{madry2017towards} and Uesato et al. \cite{uesato2018adversarial} have formulated \textit{adversarial risk} to define adversarial robustness formally, which continues to be used in this paper. In \cite{uesato2018adversarial}, Uesato et al. also analyzed that the existence of evaluation error is due to a specific attack being unable to find the worst-case adversary. In this paper, we argue the reasons causing evaluation error are more than that and try to provide a more comprehensive analysis.

First, an interesting phenomenon, named gradient traps, found in the process of evaluating robustness is introduced, which induce gradient-based attacks to fall to generate incompetent adversaries. Different from gradient masking or obfuscated gradients \cite{papernot2017practical, athalye2018obfuscated}, gradient traps do not directly lead to unusable gradients, but gradually guide the attack optimizing into traps, where the output category of models stays unchanged. In fact, we find that this phenomenon is a manifestation of evaluation error.

Then, the adversarial robustness is formulated and analyzed, from the comprehensive but intractable true risk to the tractable but incomprehensive evaluated one. Our analysis reveals that the evaluation error is composed of three components caused by compromises made in the process of making the adversarial risk computable: (1) $E_1$, caused by approximating the perceptually similar set with a computationally feasible set, (2) $E_2$, caused by replacing the non-convex and discontinuous 0-1 loss with an inappropriate surrogate loss, and (3) $E_3$, caused by employing a specific method to solve the non-convex maximization problem in a high-dimensional space. These error components may have been utilized intentionally or unintentionally by defense methods. The aforementioned gradient traps are the embodiment of $E_2$, and gradient masking reflects an extreme situation of $E_3$.

Moreover, based on the above analyses, some suggestions about adversarial robustness evaluation are provided. The principle is straightforward: an effective evaluation should diminish the error between the ideal risk and the practical one as much as possible. Experiments on CIFAR-10 \cite{krizhevsky2010convolutional} with the most empirically resistant adversarial training \cite{madry2017towards} and its variations \cite{zhang2019theoretically, wang2019bilateral, balaji2019instance, cheng2020cat} demonstrate that the error does exist. Besides, the results also indicate that no candidate can pass our more rigorous evaluation: their adversarial accuracy has dropped significantly. How to design an adversarially robust deep model is far from being solved and needs more indepth research.

\section{Related Works}
\subsection{Adversarial Examples}
By performing almost imperceptible changes to clean input data, adversarial examples can mislead deep models with high confidence. Such examples are first noticed by Szegedy et al. \cite{szegedy2013intriguing} and have attracted a lot of attention. Goodfellow et al. \cite{goodfellow2014explaining} believed that the existence of adversarial examples is because of the linear nature of neural networks, and proposed a single-step gradient ascent attack named fast gradient sign method (FGSM) to construct adversaries. Subsequently, some works \cite{kurakin2016adversarial, madry2017towards, dong2018boosting} have been proposed to extend the single-step attack to multi-step iterations. Among them, project gradient descent (PGD) proposed by Madry et al. \cite{madry2017towards} is the most typical one and shows a strong attack capability. Carlini and Wagner \cite{carlini2017towards} developed a class of powerful target attacks (C\&W attacks) by transforming the constrained optimization problem to the unconstrained one with a penalty. Su et al. \cite{su2019one} designed the One-Pixel attack and demonstrated that deep models can be fooled by only modifying one pixel in the input image. Other works \cite{xiao2018generating, song2018constructing} proposed to utilize generative models to construct adversaries.

In constructing adversarial examples, the primary criterion is that the instances with or without perturbations should perceive almost the same or similar to humans. The aforementioned methods mostly conform to this principle by limiting the $l_p$-norm of the difference between the clean data and its corresponding adversary to a small value. Besides, there are some works that maintain high perceptual similarity without the $l_p$-norm limit, and these methods can refer to as unrestricted adversarial examples \cite{engstrom2017rotation, brown2017Adversarial, sharif2017adversarial, xiao2018spatially, alaifari2018adef, song2018constructing, brown2018unrestricted}. For example, Engstrom et al. \cite{engstrom2017rotation} found that neural networks are vulnerable to simple image transformations, such as rotation and translation. Xiao et al. \cite{xiao2018spatially} proposed to construct adversaries by slightly flowing the position of each pixel in images.

\subsection{Adversarial Training}
Many defenses have been come into the scene to improve the adversarial robustness of deep models, such as data preprocessing \cite{guo2017countering, xie2017mitigating, kou2019enhancing}, feature squeezing \cite{xu2017feature}, model ensemble \cite{tramer2017ensemble, sen2020empir} and certified defenses \cite{raghunathan2018certified}. Among them, the most intuitive and empirically effective one is reusing adversarial examples as a kind of data augmentation to train deep neural networks, i.e., adversarial training (AT). These adversarial training methods are first introduced by Goodfellow et al. \cite{goodfellow2014explaining}. Madry et al. \cite{madry2017towards} developed it by training deep models with stronger adversaries generated by PGD. Subsequent works mainly focus on accelerating training \cite{shafahi2019adversarial, zhang2019you} and improving the resistance \cite{tramer2017ensemble, tramer2019adversarial, wang2019bilateral, wang2019improving}. Zhang et al. \cite{zhang2019theoretically} characterized the trade-off between accuracy and robustness and proposed TRADES, which optimizes a regularized surrogate loss to improve adversarial robustness of deep models. Wang and Zhang \cite{wang2019bilateral} proposed bilateral adversarial training (BAT), where both the data and the label are perturbed. Balaji et al. \cite{balaji2019instance} argued that adopting a uniform perturbed budget around every training data sample in standard adversarial training produces poor decision boundaries, and proposed an adaptive method named instance adaptive adversarial training (IAAT). Furthermore, Cheng et al. \cite{cheng2020cat} assigned customized labels to adversaries with label smoothing technique \cite{szegedy2016rethinking} and proposed the customized adversarial training (CAT). Tramer and Boneh \cite{tramer2019adversarial} developed multiple perturbation adversarial training (MPAT) to defense multiple types of attacks simultaneously.

\subsection{Adversarial Robustness Evaluation}
Effectively evaluating the adversarial robustness of deep models is an important issue for both academic research and practical applications, but yet still unclear. Most of previous works evaluated their defense by testing on few attack methods, usually FGSM \cite{goodfellow2014explaining}, PGD \cite{madry2017towards}, DeepFool \cite{moosavi2016deepfool} or C\&W attacks \cite{carlini2017towards}. However, such direct evaluation is sometimes insufficient. Athalye et al. \cite{athalye2018obfuscated} identified obfuscated gradients, which are a common occurrence in previous defenses, that cause gradient-based attacks \cite{goodfellow2014explaining, madry2017towards} to fail to construct offensive adversaries and lead to overestimation of the robustness of models. Such ``off-site'' factors, e.g., gradient masking and obfuscated gradients, make some specific attacks to be inefficient, rather than eliminating adversarial examples around data points fundamentally. Recently, Tramèr et al. \cite{tramer2020adaptive} studied thirteen defense methods published at top AI conferences and the results were frustrating: these defenses can be broken by properly performing adaptive attacks. These results call for more emphasis on adversarial robustness evaluation. Hendrycks and Dietterich \cite{hendrycks2019benchmarking} proposed two datasets to benchmarking DNNs robustness to common corruptions and perturbations, which cannot reflect the adversarially worst cases. Currently, one of the most effective evaluation procedures is proposed Carlini et al. \cite{carlini2019evaluating}, where a specific checklist is suggested. One concern is that this procedure mainly relies on human experience to design, and the doubt that can the results represent the true adversarial robustness still exists.

\section{Gradient Traps}
Before analyzing the evaluation error, we introduce an interesting and related observation named gradient traps. Considering the following question: under an ideal optimization condition, when an attack method attacks an adversarially trained model, why some adversaries fail to attack while others succeed. Naturally, the answer we conceived is that robust learning methods bring different levels of robustness to the model: outputs over some points are insensitive to adversarial variations of inputs and the others are not.

To verify the answer, some experiments on CIFAR-10 \cite{krizhevsky2010convolutional} with VGG-16 \cite{simonyan2014very} are conducted. Given a deep model $f$ trained with adversarial training \cite{madry2017towards}, we use PGD \cite{madry2017towards} to test $f$, and an adversary can be generated by iteratively calculating: 
\begin{equation}
x^{k+1} = \Pi(x^k + \alpha \cdot \operatorname{sgn}(\nabla_{x^k} \ell(f(x^k), y))) \text{,}
\end{equation}
where x, y denote the input data and its label respectively, $\operatorname{sgn}$ denotes the sign function, $\ell$ denotes a loss function (cross-entropy), $\alpha = \epsilon / K$ determines the step size, $\epsilon$ and $K$ are the perturbed budget and the total number of iterations respectively, $\Pi$ projects $x^{k+1}$ and the perturbation to predefined ranges, and the adversary $x' = x^K$. 

\begin{figure}[h]
\centering
\includegraphics[width=8.0cm]{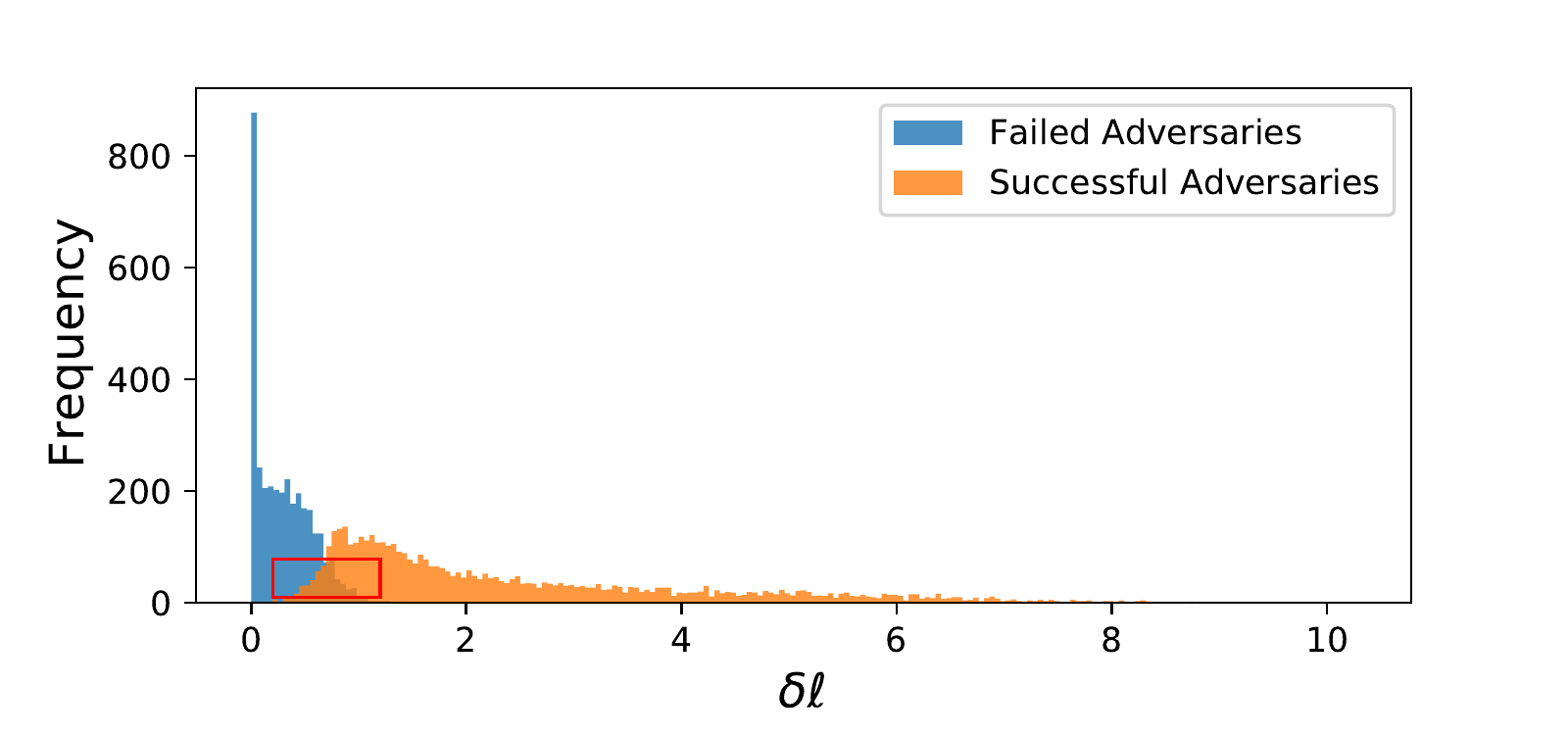}
\caption{Frequency histogram of $\delta\ell$ over failed adversaries and successful adversaries on CIFAR-10 with adversarially trained VGG-16. The position marked by the red rectangle represents the overlapping area.}
\label{fig:traps_freq_hist}
\end{figure}

To quantify the variations of the model output in adversarial setting, we define 
\begin{equation}
\delta\ell = \ell(f(x'),y) - \ell(f(x),y) \text{,}
\end{equation}
and count $\delta\ell$ over failed adversaries and successful adversaries, the results are shown in Figure \ref{fig:traps_freq_hist}.

Consistent with our conceived answer, on the whole, the successful adversaries can get larger variations of loss than the failed ones. However, we also find that there is an overlapping area, as marked with the red rectangle in Figure \ref{fig:traps_freq_hist}, where some failed samples have a similar or even larger $\delta\ell$ as the successful ones. It indicates that $\delta\ell$ is not the only factor in determining the failure of an adversary. 

\begin{figure}[h]
\centering
\includegraphics[width=8.0cm]{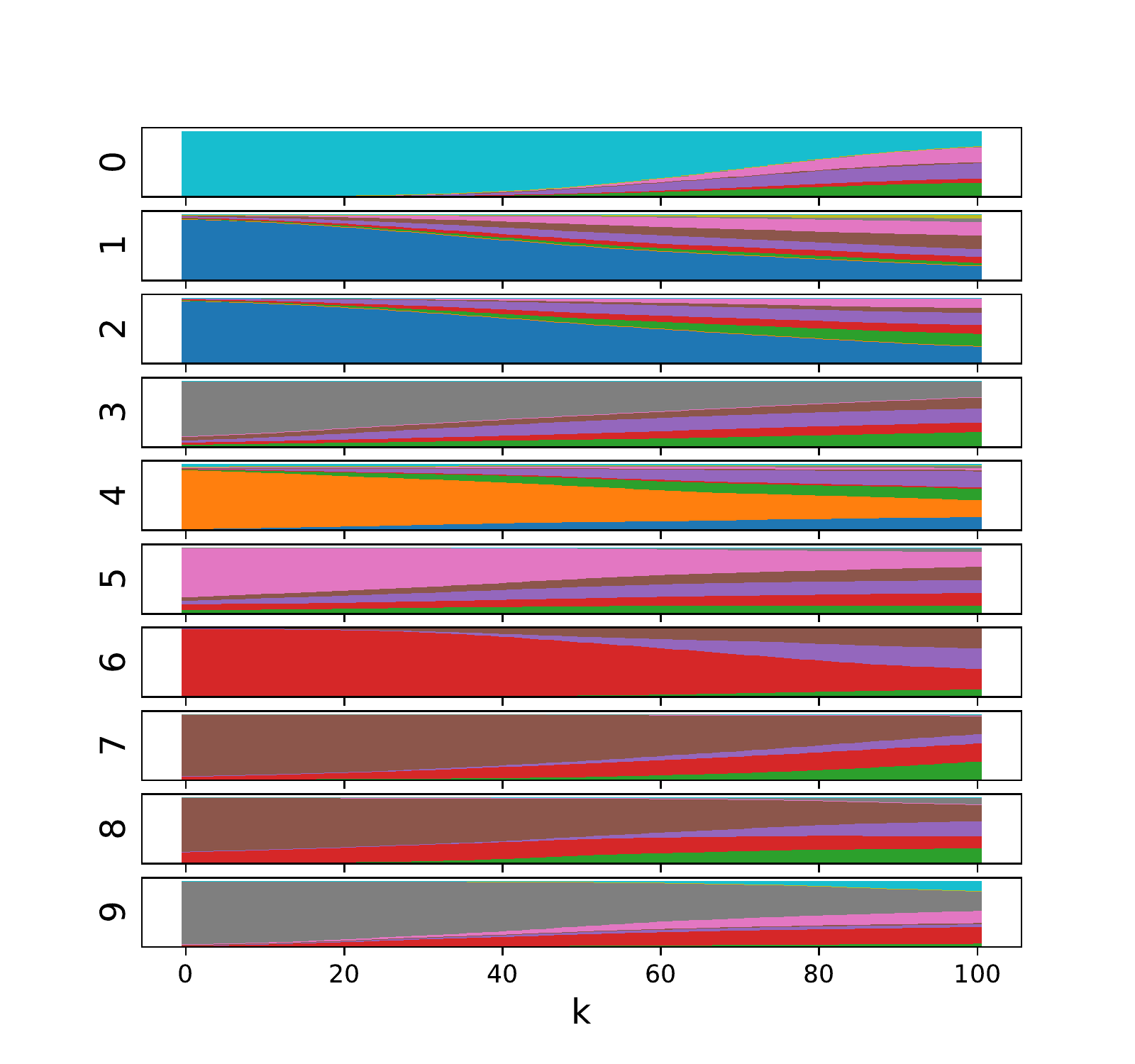}
\caption{Output confidences of the failed adversaries with 10 maximum $\delta\ell$, where the abscissa represents the progress of the attack. Each color represents the confidence of a category. $k$ denotes the number of current iterations.}
\label{fig:traps_confs}
\end{figure}

So what else? To make it clear, we select the failed adversaries with the largest 10 $\delta\ell$ and observe the changes in output confidences from clean samples to adversaries, as shown in Figure \ref{fig:traps_confs}, where each color represents the confidence of a category. The results is obvious: there is no dominant category that appeared in these failed adversaries, the allocations are too scattered. We call this phenomenon as gradient traps. Moreover, we will later show that gradient traps are one of the manifestations of evaluation error.

The above experiments indicate that, even if there are no obfuscated gradients or gradient masking, the reason why some adversaries failed is not simply because of robustness. Therefore, we must clearly understand the gap between the specific evaluated robustness and the real one.

\section{Analysis of Evaluation Error}
In this part, we try to analyze the error between the evaluated robustness and the true one. Following Uesato et al. \cite{uesato2018adversarial} and Madry et al. \cite{madry2017towards}, we first view a more general form of adversarial risk and use it to define adversarial robustness:
\begin{equation}
R_1(f) = \mathop{\mathbb{E}}\limits_{(x,y) \sim D} \bigg[\max_{x' \in S(x)} \ell_{01} (f(x'), y) \bigg] \text{,}
\label{equ:adv_risk}
\end{equation}
where $(x,y) \sim D$ denote the input data and its ground-truth label pair sampled from a joint distribution $D$, $f$ denotes the trained deep model, and $S(x)$ denotes the set of points perceptually almost same or similar to $x$, $\ell_{01}$ denotes the 0-1 loss, i.e., $\ell_{01}(a, b) = 0$ if $a = b$, $\ell_{01}(a, b) = 1$ otherwise. Adversarial risk provides a certificate of models' worst-case robustness at a fixed set of points.

However, although it reflects the true adversarial robustness, accurately computing $R_1(f)$ is impractical and impossible. Three compromises on some conditions have to be made to get an approximate estimate of $R_1(f)$. Unfortunately, such processes introduces three components of evaluation error, where two of them are inevitable and the other may be eliminated by careful choice. We formulate the error step by step according to the specific concession.

The first one is about the allowed set of adversarial examples. As described in the definition \cite{szegedy2013intriguing, goodfellow2014explaining}, an adversary is generated by perturbing the original input with (almost) imperceptible changes, so it should perceptually similar to human with the clean one. Naturally, as depicted in Equation \ref{equ:adv_risk}, the ideal allowable set of $x'$ is $S(x)$. However, $S(x)$ is a fuzzy, unclear set based on human perception, the calculation of the risk cannot be proceeded. Current, the most typical approximation is to compute in a certain range, e.g, inside a $l_p$-norm ball, i.e., 
\begin{equation}
R_2(f) = \mathop{\mathbb{E}}\limits_{(x,y) \sim D} \bigg[\max_{x' \in B(x, \epsilon)} \ell_{01} (f(x'), y) \bigg] \text{,}
\end{equation}
where $B(x,\epsilon)$ denotes a certain set where the range is controlled by a hyperparameter $\epsilon$. 

\begin{figure}[h]
\centering
\includegraphics[width=8.0cm]{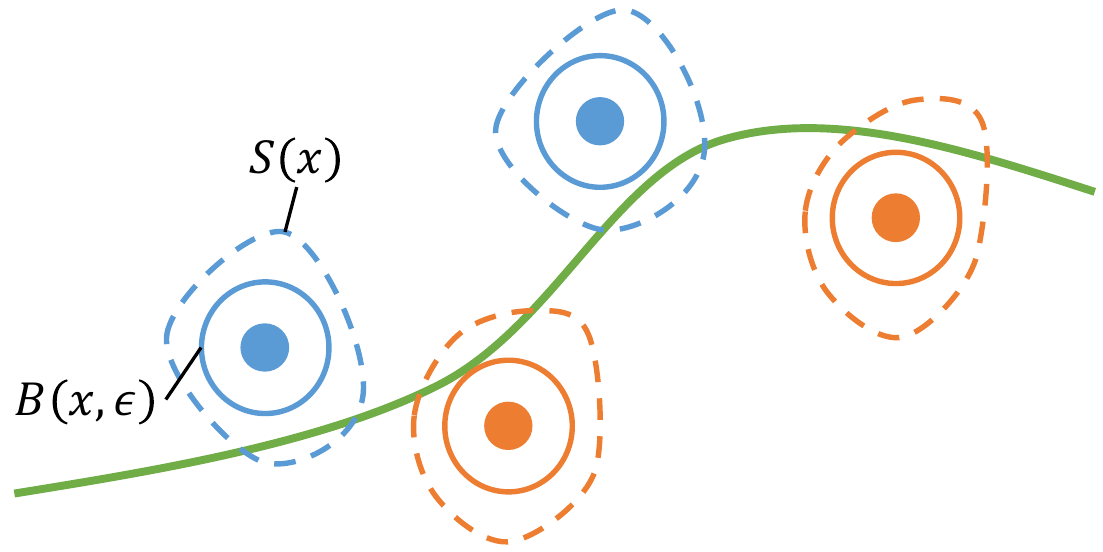}
\caption{An example of a binary classifier and its decision boundary to illustrate $E_1(f)$, where $R_2(f) = 0$ means the most robust while the true risk $R_1(f) = 1$ means the least robust, two opposite extremes.}
\label{fig:error_1}
\end{figure}

In general, $B(x,\epsilon)$ is a proper subset of $S(x)$, i.e., $B(x,\epsilon) \subset S(x)$, so $R_2(f)$ underestimates the true risk, i.e., $R_2(f) < R_1(f)$. We argue that the gap between $R_1(f)$ and $R_2(f)$ should be taken seriously, because in practical applications, hackers will not have any burden to use adversaries out of $B(x,\epsilon)$, as long as they are offensive and concealed. Moreover, the gap could be very large, as an extreme example is shown in Figure \ref{fig:error_1}, where $R_2(f) = 0$ means the most robust while $R_1(f) = 1$ means the least robust. We define $E_1(f) = R_1(f) - R_2(f)$ to formulate the gap.

The next compromise is about the loss function. As is well known, directly computing the maximum of the non-convex and discontinuous 0-1 loss is computationally intractable \cite{arora1997hardness, bartlett2006convexity}. So the common practice is to maximize a convex surrogate loss function, e.g., cross-entropy loss, that is: 
\begin{equation}
\begin{split}
R_3(f) = \mathop{\mathbb{E}}\limits_{(x,y) \sim D} \bigg[\ell_{01} (f(x'), y) \bigg] & \\
\text{s.t.\qquad} x' = \mathop{\arg\max}\limits_{x' \in B(x, \epsilon)} \ell_{s} (f(x'), y) &
\end{split} \text{,}
\end{equation}
where $\ell_{s}$ denotes a surrogate loss. 

An important but neglected issue is that can the point $x'$ that maximize $\ell_{s}$ also maximize $\ell_{01}$? If the answer is positive, then there is no gap between $R_3(f)$ and $R_2(f)$, i.e., $R_3(f) = R_2(f)$, otherwise $R_3(f) < R_2(f)$. It worth noting that, a similar but different question named surrogate risk consistency has been studied in empirical risk minimization \cite{bartlett2006convexity, tewari2007consistency}. Theoretically, Bartlett et al. \cite{bartlett2006convexity} and Tewari et al. \cite{tewari2007consistency} have proved that the optimization is consistent for most surrogate losses for binary-class and multi-class respectively. However, the conclusion of optimizing consistency cannot be directly transferred to the adversarial risk, because in adversarial setting, the goal is to find an adversary $x'$ within $B(x, \epsilon)$ to maximize $\ell_{s}$, rather than find optimal weights of $f$ to minimize $\ell_{s}$ in empirical risk setting.

On the contrary, we argue that for multi-class, inappropriate choice of surrogate loss makes the consistency cannot be guaranteed, namely, $x'$ that maximizes an inappropriate $\ell_{s}$ may not maximize $\ell_{01}$. We use a toy example to illustrate it.

\begin{figure}[h]
\centering
\includegraphics[width=8.0cm]{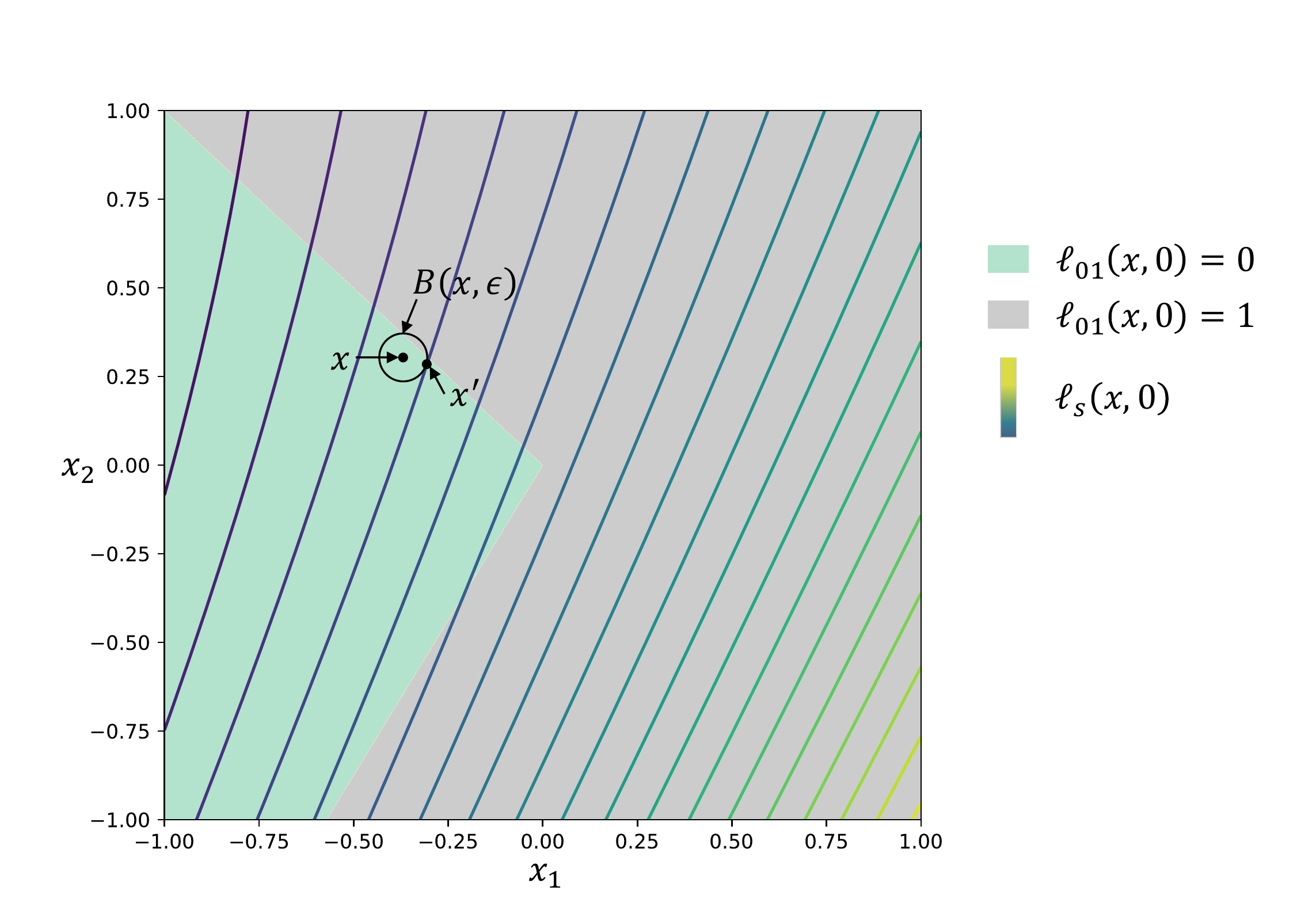}
\caption{An example of a 3-class single-layer neural network to illustrate optimization inconsistency between the 0-1 loss and the surrogate loss in adversarial settings, where $x'$ maximizes $\ell_{s}$ while $\ell_{01} = 0$.}
\label{fig:error_2}
\end{figure}

Considering a 3-class problem. Given an input $x \in [-1, 1]^2$, supposing a single-layer neural network with softmax is used to classify $x$ into three categories $y \in \{0, 1, 2\}$, i.e., 
\begin{equation}
f(x) = \operatorname{softmax}\bigg(\begin{bmatrix} 0.2 & 0.8 \\ 0.9 & 0.4 \\ 0.3 & 0.9 \end{bmatrix} \cdot \begin{bmatrix} x_0 \\ x_1 \end{bmatrix} \bigg) \text{,}
\end{equation}
and the prediction  $p = \arg\max f(x)$. What we are interested in is the values of the 0-1 loss and the values of the surrogate loss. Specifically, we adopt the cross-entropy loss as the proxy and calculate the values of $\ell_{01}(x, 0)$ and $\ell_{s}(x, 0)$. The results are shown in Figure \ref{fig:error_2}. Apparently, within $B(x, \epsilon)$, $x'$ that maximize $\ell_{s}$ may not maximize $\ell_{01}$. This error component caused by the inconsistency is defined as $E_2(f) = R_2(f) - R_3(f)$.

Directly computing $R_3(f)$ is still difficult, because a deep neural network $f$ is a highly nonlinear non-convex function with high-dimensional data as input, it is usually impossible to get the inner global maximum. We, therefore, have to compromise further: using gradient ascent or other methods to find an approximately optimal solution, that is,
\begin{equation}
\begin{split}
R_4(f) = \mathop{\mathbb{E}}\limits_{(x,y) \sim D} \bigg[\ell_{01} (f(x'), y) \bigg] & \\
\text{s.t.\qquad} x' = M(x, y, f, \ell_{s}, B, \epsilon) &
\end{split} \text{,}
\end{equation}
where $M$ denotes a specific method such as PGD \cite{madry2017towards, kurakin2016adversarial} or ES \cite{rechenberg1978evolutionsstrategien}. This process brings the third part of evaluation error, i.e., $E_3(f) = R_3(f) - R_4(f)$. It worth noting that Uesato et al. \cite{uesato2018adversarial} defined a similar one to $E_3$ named \textit{obscurity to an adversary}.

From the above formalized analyses, it can be seen that there are three compromises between the true adversarial risk $R_1(f)$ and the practical one $R_4(f)$. Every compromise is made for a more feasible computation, but it introduces an estimation error component. We summarize the error as follows.

The first one, $E_1(f)$, is caused by approximating the perceptually similar set $S(x)$ with a computationally feasible set $B(x, \epsilon)$. Sometimes this component will be so large that completely misleading to a wrong estimate of adversarial robustness, as shown in Figure \ref{fig:error_1}. This may be the hardest part to deal with, so more attention should be paid to. The next one, $E_2(f)$, is caused by approximating the non-convex and discontinuous 0-1 loss $\ell_{01}$ with an inappropriate surrogate loss $\ell_{s}$. The aforementioned gradient traps are the specific embodiment of this error component, where the change of cross-entropy value does not bring about the change of category. Besides, $E_2(f)$ may be the only one of the three that can be completely eliminated. The last one, $E_3(f)$, is caused by employing a specific method $M$ to solve the non-convex problem in a high-dimensional space. Gradient masking or obfuscated gradients that lead to incorrectly evaluation is the extreme manifestation of this part: the specific method $M$ loses the ability to find an aggressive adversary.

The above formulations also provide the principle of evaluating adversarial robustness: an effective procedure should be designed to diminish the error between the practical risk and the truth risk as much as possible. 

\section{Evaluation Suggestions}
Following the aforementioned principle we give our suggestions about adversarial robustness evaluation. Unlike the excellent and detailed recommendations presented by Carlini et al. \cite{carlini2019evaluating}, our suggestions are instructive. The suggestions are as follows.

\noindent \textbf{Considering multiple types of attacks.} Despite a lot of works study a threat model with $l_p$-bounded intensity perturbations, which have well-designed nature, we argue that evaluating against more types of attacks, like location perturbations \cite{engstrom2017rotation, xiao2018generating, alaifari2018adef}, are more meaningful to the practical deployment of deep neural networks. Considering multiple approximate sets, i.e., $\mathcal{B} = \{B_1, B_2, ..., B_n\}$, is more likely to be close to the true similar set $S$.

\noindent \textbf{Using various surrogate losses.} Currently, the cross-entropy loss is the most common surrogate. However, for multi-class tasks, maximize the surrogate loss may not maximize the 0-1 loss, which leads to overestimating the adversarial robustness of models. Although we have found that some surrogates may completely eliminate this gap, we recommend to use various losses, i.e., $\mathcal{L} = \{\ell_{1}, \ell_{2}, ..., \ell_{m}\}$, rather than a single one before formal proofs.

\noindent \textbf{Solving the maximum with diverse methods.} When solving a high-dimensional nonlinear problem, using a single method often fall into a local optimum. Moreover, this method may appear in the training process, leading to the trained model overfit to it. Therefore, an evaluation should solve the maximum problem with multiple methods $\mathcal{M} = \{M_1, M_2, ..., M_k\}$.

\noindent \textbf{Considering the worst case.} Here is the key: what we want to evaluate is the adversarial robustness of a model, rather than its defensive ability against multiple attacks. Thus, over multiple types of attacks, surrogates and methods, we consider these as a whole and focus on the worst case. Namely, a deep model $f$ can be identified as adversarial robust around a point $x$ if and only if it passes all $n \times m \times k$ tests. Although very strict, we think this is the best way so far to make the evaluated risk close to the true one.  

\section{Experiments}
The purposes of our experiments are two folds: (1) verify whether the error exists empirically, and (2) evaluate the adversarial robustness of existing methods under our strict conditions.

\subsection{Setup}
All experiments are conducted on CIFAR-10 \cite{krizhevsky2010convolutional} with VGG-16 \cite{simonyan2014very}. Due to the relatively small size of CIFAR images, the first convolution layer is replaced with 3$\times$3 kernel size, and the final classification layers are replaced by global average pooling \cite{lin2013network} + single fully connection layer. All models are trained using SGD optimizer with momentum of 0.9 and weight decay of 5e-4. The batch size is 256 and the total training duration is 80. The initial learning rate is set to 0.1 and dropped by 10 after 30 and 60 epochs. We employ random cropping and random horizontal flipping with probability set to 0.5 as data augmentations and all images are normalized to [0-1]. Our experiments are implemented with PyTorch \cite{paszke2017automatic} on a GeForce RTX 2080 Ti GPU.

\subsection{Evaluated Defenses}
Defense methods considered in this paper including adversarial training (AT) \cite{madry2017towards}, and its variations, including TRADES \cite{zhang2019theoretically}, BAT \cite{wang2019bilateral}, IAAT \cite{balaji2019instance} and CAT \cite{cheng2020cat}. The reason for choosing adversarial training as the experimental object is that it is empirically the most effective method so far, and obfuscated gradients do not appear in it \cite{athalye2018obfuscated}. All hyperparameters are set according to the original literature.

\subsection{Evaluation Details}
We detail the evaluation adopted in our experiments. It worth noting that our evaluation is mainly designed based on the selected defenses, i.e., adversarial training and its variations, and is not necessarily applicable to other methods. Readers can appropriately consider our suggestions and design their own evaluations.

\begin{figure}[h]
\centering
\includegraphics[width=8.0cm]{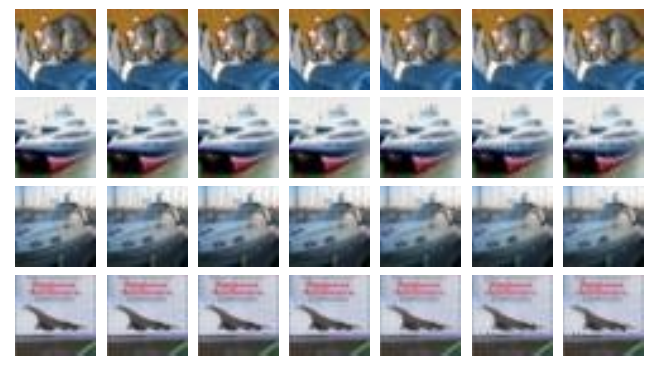}
\caption{Examples of adversaries under different sets. From left to right: clean, $(I, l_{\infty}, \epsilon=0.024)$, $(I, l_{1}, \epsilon=30.5)$, $(I, l_{2}, \epsilon=0.78)$, $(L, l_{\infty}, \epsilon=0.019)$, $(L, l_{1}, \epsilon=8.55)$, $(L, l_{2}, \epsilon=0.38)$.}
\label{fig:sets}
\end{figure}

First of all, for the selection of allowed sets $\mathcal{B}$, we consider two major types of attacks, intensity perturbations ($I$) \cite{goodfellow2014explaining, madry2017towards, carlini2017towards} and location perturbations ($L$) \cite{xiao2018spatially, alaifari2018adef}. For each type, we involve three bounds, $l_{\infty}$, $l_{1}$, and $l_{2}$, and therefore there are six sets in total. To determine the perturbed budget $\epsilon$ for each set, we use binary search increasing from 0 and choose the one that first reducing the accuracy of a standard trained model (without any defense) to less than 0.01. The allowed sets adopted in this paper is $\mathcal{B} = \{(I, l_{\infty}, \epsilon=0.024), (I, l_1, \epsilon=30.5), (I, l_2, \epsilon=0.78), (L, l_{\infty}, \epsilon=0.019), (L, l_1, \epsilon=8.55), (L, l_2, \epsilon=0.38)\}$. Some examples of adversaries are shown in Figure \ref{fig:sets} to illustrate that the selection of allowed sets do not violate the principle of perceived similarity.

Secondly, four types of surrogate losses are used in our evaluation, including cross-entropy loss $CE(f(x), y)$ (CE), negative target cross-entropy loss $-CE(f(x), t)$ (tCE), logistic loss $\max_{i \neq y}z(x)_i - z(x)_y$ (LL), and target logistic loss $z(x)_t - z(x)_y$ (tLL), where $z(x)$ denotes the logits, i.e., the output of all layers except the softmax. So the selected surrogate losses are $\mathcal{L} = \{CE(f(x), y), -CE(f(x), 0), -CE(f(x), 1), ..., \max_{i \neq y}z(x)_i - z(x)_y, z(x)_{0} - z(x)_{y}, z(x)_{1} - z(x)_{y}, ...\}$, where the total number is $22$ for CIFAR-10 dataset.

Next, the selection of search methods $\mathcal{M}$. Due to the subjects, adversarial training and its four variants, have not been found to rely on obfuscated gradients, we use the naive project gradient descent (ascent) without any pertinent technique such as Expectation Over Transformation \cite{athalye2018synthesizing}. For each adversarially trained model, we examine it with one white-box attack and four transfer-based attacks, which together constitute $\mathcal{M}$. 

\begin{table*}[t]
\centering
\caption{$A_{me}$, $A_{mi}$, and $A_{wc}$ over different states on CIFAR-10 with adversarial trained models.}
\begin{tabular}{l|l|c|c|c|c|c|c|c}
\toprule
						&          & Clean & S0    & S1    & S2    & S3    & S4    & S5    \\ \midrule
\multirow{3}{*}{AT} 	& $A_{me}$ & \multirow{3}{*}{0.818} & 0.623 & 0.701 & 0.776 & 0.732 & 0.727 & 0.692 \\
						& $A_{mi}$ &   & 0.623 & 0.623 & 0.609 & 0.419 & 0.419 & 0.255 \\
						& $A_{wc}$ &   & 0.623 & 0.619 & 0.565 & 0.035 & 0.035 & 0.009 \\ \midrule
\multirow{3}{*}{TRADES} & $A_{me}$ & \multirow{3}{*}{0.875} & 0.622 & 0.727 & 0.818 & 0.764 & 0.759 & 0.716 \\
						& $A_{mi}$ &   & 0.622 & 0.622 & 0.621 & 0.404 & 0.404 & 0.197 \\
						& $A_{wc}$ &   & 0.622 & 0.622 & 0.564 & 0.036 & 0.036 & 0.009 \\ \midrule
\multirow{3}{*}{BAT}    & $A_{me}$ & \multirow{3}{*}{0.822} & 0.617 & 0.703 & 0.776 & 0.731 & 0.726 & 0.692 \\
						& $A_{mi}$ &   & 0.617 & 0.617 & 0.607 & 0.425 & 0.425 & 0.285 \\
						& $A_{wc}$ &   & 0.617 & 0.616 & 0.558 & 0.012 & 0.012 & 0.004 \\ \midrule
\multirow{3}{*}{IAAT}   & $A_{me}$ & \multirow{3}{*}{0.871} & 0.630 & 0.726 & 0.818 & 0.768 & 0.762 & 0.721 \\
						& $A_{mi}$ &   & 0.630 & 0.630 & 0.630 & 0.426 & 0.426 & 0.228 \\
						& $A_{wc}$ &   & 0.630 & 0.629 & 0.578 & 0.021 & 0.021 & 0.006 \\ \midrule
\multirow{3}{*}{CAT}    & $A_{me}$ & \multirow{3}{*}{0.884} & 0.627 & 0.732 & 0.823 & 0.772 & 0.767 & 0.724 \\
						& $A_{mi}$ &   & 0.627 & 0.627 & 0.613 & 0.437 & 0.437 & 0.307 \\
						& $A_{wc}$ &   & 0.627 & 0.625 & 0.564 & 0.007 & 0.007 & 0.003 \\ \midrule
\end{tabular} \label{tab:me_mi_wc}
\end{table*} 

Last but not least, in order to remain the same as the previous literature, we report adversarial accuracy as the evaluation index:
\begin{equation}
\begin{split}
A = 1 - R_4(f) = 1 - \mathop{\mathbb{E}}\limits_{(x,y) \sim D} \bigg[\ell_{01} (f(x'), y) \bigg] & \\
\text{s.t.\qquad} x' = M(x, y, f, \ell_{s}, B, \epsilon) &
\end{split} \text{,}
\end{equation}
where $A = 0$ means the least robust and $A = 1$ means the most robust.

\subsection{Results}
As depicted above, each adversarially trained model needs to be tested $6 \times 22 \times 5 = 660$ times, and the detailed results of AT, TRADES, BAT, IAAT, and CAT are shown in the supplementary material. In order to show the results more intuitively, we define some states and calculate the mean accuracy ($A_{me}$), minimum accuracy ($A_{mi}$), and worse-case accuracy ($A_{wc}$) in that state, as shown in Table \ref{tab:me_mi_wc}. For clarity, the worst case used here means that a deep model is robust around an instance $x$ if and only if all of 660 adversarial versions are classified correctly. The states include:
\begin{itemize}
\item S0, considering the white-box intensity-based $l_{\infty}$-bounded attack with the cross-entropy loss.
\item S1, considering all intensity-based $l_{\infty}$-bounded attacks with the cross-entropy loss.
\item S2, considering all intensity-based $l_{\infty}$-bounded attacks with all surrogate losses.
\item S3, considering all intensity-based $l_{\infty}$-bounded, $l_{1}$-bounded attacks with all surrogate losses.
\item S4, considering all intensity-based attacks with all surrogate losses.
\item S5, considering all intensity-based and location-based attacks with all surrogate losses.
\end{itemize}

Ignoring the worst-case scenario, all five defense methods can improve the robustness of deep models to a certain extent. For frequently used white-box intensity-based $l_{\infty}$-bounded PGD attack, TRADES, IAAT, CAT have achieve relatively good results in terms of balancing clean accuracy (0.875, 0.871, 0.884) and adversarial accuracy (0.622, 0.630, 0.627). Considering Liebig's law of the minimum, over all cases, CAT has the relatively largest minimum of adversarial accuracy (0.307). The average adversarial accuracy of all defenses is greater than 0.690.

These results seem to indicate that the selected defenses, especially CAT, can achieve good resistance against adversarial examples. However, after the aforementioned analyses, we argue that considering the worse case is helpful to reduce the error between evaluated risk and truth risk, and it is more in line with the practical situation. Unfortunately, in this case, the adversarial accuracy of all defenses drops to 0. Even worse, what seemed to be the most robust method, CAT, performs the worst (0.003).

Obviously in the table, the declines in adversarial accuracy mainly occur in two stages: from S1 to S2 and from S2 to S3. The decay of accuracy from S1 to S2 means that in adversarial setting, the problem of inconsistent optimization between the surrogate loss and the 0-1 loss does exist, not only for toy sets. 

The largest part is the attenuation from S2 to S3, it indicates that we should get out of the ``comfort zone'', which limited to a single bound, and consider more in line with the actual situation. Interestingly, the location perturbations we prepared for the defenses are not used, because the methods do not need to wait for these attacks to be broken.

The accuracy decay between S0 and S1 is relatively small, and we think there are two reasons: (1) the selected defenses do not use gradient masking or obfuscated gradients to invalidate PGD, and (2) the optimizer we considered is not enough, maybe some heuristic methods are needed.

\section{Conclusion}
In this paper, we introduce gradient traps and analyze the three component of evaluation error between the evaluated adversarial risk in practice and the ideal adversarial risk: (1) $E_1$, caused by approximating the perceptually similar set with a specific set, (2) $E_2$, caused by replacing the non-convex and discontinuous 0-1 loss with a surrogate loss, and (3) $E_3$, caused by employing a specific method to solve the maximum of the non-convex problem. Guided by the above analyses, we present our suggestions about how to evaluate adversarial robustness effectively, and the main principle is to reduce the error. Through conducting experiments on SOTA adversarial training and its variations, our rigorous evaluation has greatly reduced the adversarial accuracy of the defense models, from 0.623, 0.622, 0.617, 0.630, 0.627 to 0.009, 0.009, 0.004, 0.006, 0.003 for AT, TRADES, BAT, IAAT and CAT respectively. The results demonstrate the existence of the error empirically, and also shows the limitations of the previous defense methods. We hope our work will help build stronger and broader defense methods.

\section*{Acknowledgements}
The work is partially supported by the National Natural Science Foundation of China under grand No.U19B2044 and No.61836011.

\bibliographystyle{ACM-Reference-Format}
\bibliography{./references.bib}


\begin{thebibliography}{55}


\ifx \showCODEN    \undefined \def \showCODEN     #1{\unskip}     \fi
\ifx \showDOI      \undefined \def \showDOI       #1{#1}\fi
\ifx \showISBNx    \undefined \def \showISBNx     #1{\unskip}     \fi
\ifx \showISBNxiii \undefined \def \showISBNxiii  #1{\unskip}     \fi
\ifx \showISSN     \undefined \def \showISSN      #1{\unskip}     \fi
\ifx \showLCCN     \undefined \def \showLCCN      #1{\unskip}     \fi
\ifx \shownote     \undefined \def \shownote      #1{#1}          \fi
\ifx \showarticletitle \undefined \def \showarticletitle #1{#1}   \fi
\ifx \showURL      \undefined \def \showURL       {\relax}        \fi
\providecommand\bibfield[2]{#2}
\providecommand\bibinfo[2]{#2}
\providecommand\natexlab[1]{#1}
\providecommand\showeprint[2][]{arXiv:#2}

\bibitem[\protect\citeauthoryear{Alaifari, Alberti, and Gauksson}{Alaifari
  et~al\mbox{.}}{2018}]%
        {alaifari2018adef}
\bibfield{author}{\bibinfo{person}{Rima Alaifari}, \bibinfo{person}{Giovanni~S
  Alberti}, {and} \bibinfo{person}{Tandri Gauksson}.}
  \bibinfo{year}{2018}\natexlab{}.
\newblock \showarticletitle{ADef: an iterative algorithm to construct
  adversarial deformations}.
\newblock \bibinfo{journal}{\emph{arXiv preprint arXiv:1804.07729}}
  (\bibinfo{year}{2018}).
\newblock


\bibitem[\protect\citeauthoryear{Arora, Babai, Stern, and Sweedyk}{Arora
  et~al\mbox{.}}{1997}]%
        {arora1997hardness}
\bibfield{author}{\bibinfo{person}{Sanjeev Arora},
  \bibinfo{person}{L{\'a}szl{\'o} Babai}, \bibinfo{person}{Jacques Stern},
  {and} \bibinfo{person}{Z Sweedyk}.} \bibinfo{year}{1997}\natexlab{}.
\newblock \showarticletitle{The hardness of approximate optima in lattices,
  codes, and systems of linear equations}.
\newblock \bibinfo{journal}{\emph{J. Comput. System Sci.}}
  \bibinfo{volume}{54}, \bibinfo{number}{2} (\bibinfo{year}{1997}),
  \bibinfo{pages}{317--331}.
\newblock


\bibitem[\protect\citeauthoryear{Athalye, Carlini, and Wagner}{Athalye
  et~al\mbox{.}}{2018a}]%
        {athalye2018obfuscated}
\bibfield{author}{\bibinfo{person}{Anish Athalye}, \bibinfo{person}{Nicholas
  Carlini}, {and} \bibinfo{person}{David Wagner}.}
  \bibinfo{year}{2018}\natexlab{a}.
\newblock \showarticletitle{Obfuscated gradients give a false sense of
  security: Circumventing defenses to adversarial examples}.
\newblock \bibinfo{journal}{\emph{arXiv preprint arXiv:1802.00420}}
  (\bibinfo{year}{2018}).
\newblock


\bibitem[\protect\citeauthoryear{Athalye, Engstrom, Ilyas, and Kwok}{Athalye
  et~al\mbox{.}}{2018b}]%
        {athalye2018synthesizing}
\bibfield{author}{\bibinfo{person}{Anish Athalye}, \bibinfo{person}{Logan
  Engstrom}, \bibinfo{person}{Andrew Ilyas}, {and} \bibinfo{person}{Kevin
  Kwok}.} \bibinfo{year}{2018}\natexlab{b}.
\newblock \showarticletitle{Synthesizing robust adversarial examples}. In
  \bibinfo{booktitle}{\emph{International conference on machine learning}}.
  PMLR, \bibinfo{pages}{284--293}.
\newblock


\bibitem[\protect\citeauthoryear{Balaji, Goldstein, and Hoffman}{Balaji
  et~al\mbox{.}}{2019}]%
        {balaji2019instance}
\bibfield{author}{\bibinfo{person}{Yogesh Balaji}, \bibinfo{person}{Tom
  Goldstein}, {and} \bibinfo{person}{Judy Hoffman}.}
  \bibinfo{year}{2019}\natexlab{}.
\newblock \showarticletitle{Instance adaptive adversarial training: Improved
  accuracy tradeoffs in neural nets}.
\newblock \bibinfo{journal}{\emph{arXiv preprint arXiv:1910.08051}}
  (\bibinfo{year}{2019}).
\newblock


\bibitem[\protect\citeauthoryear{Bartlett, Jordan, and McAuliffe}{Bartlett
  et~al\mbox{.}}{2006}]%
        {bartlett2006convexity}
\bibfield{author}{\bibinfo{person}{Peter~L Bartlett},
  \bibinfo{person}{Michael~I Jordan}, {and} \bibinfo{person}{Jon~D McAuliffe}.}
  \bibinfo{year}{2006}\natexlab{}.
\newblock \showarticletitle{Convexity, classification, and risk bounds}.
\newblock \bibinfo{journal}{\emph{J. Amer. Statist. Assoc.}}
  \bibinfo{volume}{101}, \bibinfo{number}{473} (\bibinfo{year}{2006}),
  \bibinfo{pages}{138--156}.
\newblock


\bibitem[\protect\citeauthoryear{Brown, Carlini, Zhang, Olsson, Christiano, and
  Goodfellow}{Brown et~al\mbox{.}}{2018}]%
        {brown2018unrestricted}
\bibfield{author}{\bibinfo{person}{Tom~B Brown}, \bibinfo{person}{Nicholas
  Carlini}, \bibinfo{person}{Chiyuan Zhang}, \bibinfo{person}{Catherine
  Olsson}, \bibinfo{person}{Paul Christiano}, {and} \bibinfo{person}{Ian
  Goodfellow}.} \bibinfo{year}{2018}\natexlab{}.
\newblock \showarticletitle{Unrestricted adversarial examples}.
\newblock \bibinfo{journal}{\emph{arXiv preprint arXiv:1809.08352}}
  (\bibinfo{year}{2018}).
\newblock


\bibitem[\protect\citeauthoryear{Brown, Man{\'e}, Roy, Abadi, and Gilmer}{Brown
  et~al\mbox{.}}{2017}]%
        {brown2017Adversarial}
\bibfield{author}{\bibinfo{person}{Tom~B. Brown}, \bibinfo{person}{Dandelion
  Man{\'e}}, \bibinfo{person}{Aurko Roy}, \bibinfo{person}{Mart{\'i}n Abadi},
  {and} \bibinfo{person}{Justin Gilmer}.} \bibinfo{year}{2017}\natexlab{}.
\newblock \showarticletitle{Adversarial Patch}.
\newblock \bibinfo{journal}{\emph{arXiv preprint 1712.09665}}
  (\bibinfo{year}{2017}).
\newblock


\bibitem[\protect\citeauthoryear{Carlini, Athalye, Papernot, Brendel, Rauber,
  Tsipras, Goodfellow, Madry, and Kurakin}{Carlini et~al\mbox{.}}{2019}]%
        {carlini2019evaluating}
\bibfield{author}{\bibinfo{person}{Nicholas Carlini}, \bibinfo{person}{Anish
  Athalye}, \bibinfo{person}{Nicolas Papernot}, \bibinfo{person}{Wieland
  Brendel}, \bibinfo{person}{Jonas Rauber}, \bibinfo{person}{Dimitris Tsipras},
  \bibinfo{person}{Ian Goodfellow}, \bibinfo{person}{Aleksander Madry}, {and}
  \bibinfo{person}{Alexey Kurakin}.} \bibinfo{year}{2019}\natexlab{}.
\newblock \showarticletitle{On evaluating adversarial robustness}.
\newblock \bibinfo{journal}{\emph{arXiv preprint arXiv:1902.06705}}
  (\bibinfo{year}{2019}).
\newblock


\bibitem[\protect\citeauthoryear{Carlini and Wagner}{Carlini and
  Wagner}{2017}]%
        {carlini2017towards}
\bibfield{author}{\bibinfo{person}{Nicholas Carlini} {and}
  \bibinfo{person}{David Wagner}.} \bibinfo{year}{2017}\natexlab{}.
\newblock \showarticletitle{Towards evaluating the robustness of neural
  networks}. In \bibinfo{booktitle}{\emph{2017 ieee symposium on security and
  privacy (sp)}}. IEEE, \bibinfo{pages}{39--57}.
\newblock


\bibitem[\protect\citeauthoryear{Cheng, Lei, Chen, Dhillon, and Hsieh}{Cheng
  et~al\mbox{.}}{2020}]%
        {cheng2020cat}
\bibfield{author}{\bibinfo{person}{Minhao Cheng}, \bibinfo{person}{Qi Lei},
  \bibinfo{person}{Pin-Yu Chen}, \bibinfo{person}{Inderjit Dhillon}, {and}
  \bibinfo{person}{Cho-Jui Hsieh}.} \bibinfo{year}{2020}\natexlab{}.
\newblock \showarticletitle{Cat: Customized adversarial training for improved
  robustness}.
\newblock \bibinfo{journal}{\emph{arXiv preprint arXiv:2002.06789}}
  (\bibinfo{year}{2020}).
\newblock


\bibitem[\protect\citeauthoryear{Devlin, Chang, Lee, and Toutanova}{Devlin
  et~al\mbox{.}}{2018}]%
        {devlin2018bert}
\bibfield{author}{\bibinfo{person}{Jacob Devlin}, \bibinfo{person}{Ming-Wei
  Chang}, \bibinfo{person}{Kenton Lee}, {and} \bibinfo{person}{Kristina
  Toutanova}.} \bibinfo{year}{2018}\natexlab{}.
\newblock \showarticletitle{Bert: Pre-training of deep bidirectional
  transformers for language understanding}.
\newblock \bibinfo{journal}{\emph{arXiv preprint arXiv:1810.04805}}
  (\bibinfo{year}{2018}).
\newblock


\bibitem[\protect\citeauthoryear{Dong, Liao, Pang, Su, Zhu, Hu, and Li}{Dong
  et~al\mbox{.}}{2018}]%
        {dong2018boosting}
\bibfield{author}{\bibinfo{person}{Yinpeng Dong}, \bibinfo{person}{Fangzhou
  Liao}, \bibinfo{person}{Tianyu Pang}, \bibinfo{person}{Hang Su},
  \bibinfo{person}{Jun Zhu}, \bibinfo{person}{Xiaolin Hu}, {and}
  \bibinfo{person}{Jianguo Li}.} \bibinfo{year}{2018}\natexlab{}.
\newblock \showarticletitle{Boosting adversarial attacks with momentum}. In
  \bibinfo{booktitle}{\emph{Proceedings of the IEEE conference on computer
  vision and pattern recognition}}. \bibinfo{pages}{9185--9193}.
\newblock


\bibitem[\protect\citeauthoryear{Engstrom, Tsipras, Schmidt, and
  Madry}{Engstrom et~al\mbox{.}}{2017}]%
        {engstrom2017rotation}
\bibfield{author}{\bibinfo{person}{Logan Engstrom}, \bibinfo{person}{Dimitris
  Tsipras}, \bibinfo{person}{Ludwig Schmidt}, {and} \bibinfo{person}{Aleksander
  Madry}.} \bibinfo{year}{2017}\natexlab{}.
\newblock \showarticletitle{A rotation and a translation suffice: Fooling cnns
  with simple transformations}.
\newblock \bibinfo{journal}{\emph{arXiv preprint arXiv:1712.02779}}
  \bibinfo{volume}{1}, \bibinfo{number}{2} (\bibinfo{year}{2017}),
  \bibinfo{pages}{3}.
\newblock


\bibitem[\protect\citeauthoryear{Girshick}{Girshick}{2015}]%
        {girshick2015fast}
\bibfield{author}{\bibinfo{person}{Ross Girshick}.}
  \bibinfo{year}{2015}\natexlab{}.
\newblock \showarticletitle{Fast r-cnn}. In
  \bibinfo{booktitle}{\emph{Proceedings of the IEEE international conference on
  computer vision}}. \bibinfo{pages}{1440--1448}.
\newblock


\bibitem[\protect\citeauthoryear{Goodfellow, Shlens, and Szegedy}{Goodfellow
  et~al\mbox{.}}{2014}]%
        {goodfellow2014explaining}
\bibfield{author}{\bibinfo{person}{Ian~J Goodfellow}, \bibinfo{person}{Jonathon
  Shlens}, {and} \bibinfo{person}{Christian Szegedy}.}
  \bibinfo{year}{2014}\natexlab{}.
\newblock \showarticletitle{Explaining and harnessing adversarial examples}.
\newblock \bibinfo{journal}{\emph{arXiv preprint arXiv:1412.6572}}
  (\bibinfo{year}{2014}).
\newblock


\bibitem[\protect\citeauthoryear{Gu and Rigazio}{Gu and Rigazio}{2014}]%
        {gu2014towards}
\bibfield{author}{\bibinfo{person}{Shixiang Gu} {and} \bibinfo{person}{Luca
  Rigazio}.} \bibinfo{year}{2014}\natexlab{}.
\newblock \showarticletitle{Towards deep neural network architectures robust to
  adversarial examples}.
\newblock \bibinfo{journal}{\emph{arXiv preprint arXiv:1412.5068}}
  (\bibinfo{year}{2014}).
\newblock


\bibitem[\protect\citeauthoryear{Guo, Rana, Cisse, and Van Der~Maaten}{Guo
  et~al\mbox{.}}{2017}]%
        {guo2017countering}
\bibfield{author}{\bibinfo{person}{Chuan Guo}, \bibinfo{person}{Mayank Rana},
  \bibinfo{person}{Moustapha Cisse}, {and} \bibinfo{person}{Laurens Van
  Der~Maaten}.} \bibinfo{year}{2017}\natexlab{}.
\newblock \showarticletitle{Countering adversarial images using input
  transformations}.
\newblock \bibinfo{journal}{\emph{arXiv preprint arXiv:1711.00117}}
  (\bibinfo{year}{2017}).
\newblock


\bibitem[\protect\citeauthoryear{Heaven}{Heaven}{2019}]%
        {heaven2019deep}
\bibfield{author}{\bibinfo{person}{Douglas Heaven}.}
  \bibinfo{year}{2019}\natexlab{}.
\newblock \showarticletitle{Why deep-learning AIs are so easy to fool}.
\newblock \bibinfo{journal}{\emph{Nature}} \bibinfo{volume}{574},
  \bibinfo{number}{7777} (\bibinfo{year}{2019}), \bibinfo{pages}{163}.
\newblock


\bibitem[\protect\citeauthoryear{Hendrycks and Dietterich}{Hendrycks and
  Dietterich}{2019}]%
        {hendrycks2019benchmarking}
\bibfield{author}{\bibinfo{person}{Dan Hendrycks} {and} \bibinfo{person}{Thomas
  Dietterich}.} \bibinfo{year}{2019}\natexlab{}.
\newblock \showarticletitle{Benchmarking neural network robustness to common
  corruptions and perturbations}.
\newblock \bibinfo{journal}{\emph{arXiv preprint arXiv:1903.12261}}
  (\bibinfo{year}{2019}).
\newblock


\bibitem[\protect\citeauthoryear{Jakubovitz and Giryes}{Jakubovitz and
  Giryes}{2018}]%
        {jakubovitz2018improving}
\bibfield{author}{\bibinfo{person}{Daniel Jakubovitz} {and}
  \bibinfo{person}{Raja Giryes}.} \bibinfo{year}{2018}\natexlab{}.
\newblock \showarticletitle{Improving dnn robustness to adversarial attacks
  using jacobian regularization}. In \bibinfo{booktitle}{\emph{Proceedings of
  the European Conference on Computer Vision (ECCV)}}.
  \bibinfo{pages}{514--529}.
\newblock


\bibitem[\protect\citeauthoryear{Kou, Lee, Chang, and Ng}{Kou
  et~al\mbox{.}}{2019}]%
        {kou2019enhancing}
\bibfield{author}{\bibinfo{person}{Connie Kou}, \bibinfo{person}{Hwee~Kuan
  Lee}, \bibinfo{person}{Ee-Chien Chang}, {and} \bibinfo{person}{Teck~Khim
  Ng}.} \bibinfo{year}{2019}\natexlab{}.
\newblock \showarticletitle{Enhancing Transformation-Based Defenses Against
  Adversarial Attacks with a Distribution Classifier}. In
  \bibinfo{booktitle}{\emph{International Conference on Learning
  Representations}}.
\newblock


\bibitem[\protect\citeauthoryear{Krizhevsky and Hinton}{Krizhevsky and
  Hinton}{2010}]%
        {krizhevsky2010convolutional}
\bibfield{author}{\bibinfo{person}{Alex Krizhevsky} {and}
  \bibinfo{person}{Geoff Hinton}.} \bibinfo{year}{2010}\natexlab{}.
\newblock \showarticletitle{Convolutional deep belief networks on cifar-10}.
\newblock \bibinfo{journal}{\emph{Unpublished manuscript}}
  \bibinfo{volume}{40}, \bibinfo{number}{7} (\bibinfo{year}{2010}),
  \bibinfo{pages}{1--9}.
\newblock


\bibitem[\protect\citeauthoryear{Krizhevsky, Sutskever, and Hinton}{Krizhevsky
  et~al\mbox{.}}{2012}]%
        {krizhevsky2012imagenet}
\bibfield{author}{\bibinfo{person}{Alex Krizhevsky}, \bibinfo{person}{Ilya
  Sutskever}, {and} \bibinfo{person}{Geoffrey~E Hinton}.}
  \bibinfo{year}{2012}\natexlab{}.
\newblock \showarticletitle{Imagenet classification with deep convolutional
  neural networks}. In \bibinfo{booktitle}{\emph{Advances in neural information
  processing systems}}. \bibinfo{pages}{1097--1105}.
\newblock


\bibitem[\protect\citeauthoryear{Kurakin, Goodfellow, and Bengio}{Kurakin
  et~al\mbox{.}}{2016}]%
        {kurakin2016adversarial}
\bibfield{author}{\bibinfo{person}{Alexey Kurakin}, \bibinfo{person}{Ian
  Goodfellow}, {and} \bibinfo{person}{Samy Bengio}.}
  \bibinfo{year}{2016}\natexlab{}.
\newblock \showarticletitle{Adversarial machine learning at scale}.
\newblock \bibinfo{journal}{\emph{arXiv preprint arXiv:1611.01236}}
  (\bibinfo{year}{2016}).
\newblock


\bibitem[\protect\citeauthoryear{Lin, Chen, and Yan}{Lin et~al\mbox{.}}{2013}]%
        {lin2013network}
\bibfield{author}{\bibinfo{person}{Min Lin}, \bibinfo{person}{Qiang Chen},
  {and} \bibinfo{person}{Shuicheng Yan}.} \bibinfo{year}{2013}\natexlab{}.
\newblock \showarticletitle{Network in network}.
\newblock \bibinfo{journal}{\emph{arXiv preprint arXiv:1312.4400}}
  (\bibinfo{year}{2013}).
\newblock


\bibitem[\protect\citeauthoryear{Madry, Makelov, Schmidt, Tsipras, and
  Vladu}{Madry et~al\mbox{.}}{2017}]%
        {madry2017towards}
\bibfield{author}{\bibinfo{person}{Aleksander Madry},
  \bibinfo{person}{Aleksandar Makelov}, \bibinfo{person}{Ludwig Schmidt},
  \bibinfo{person}{Dimitris Tsipras}, {and} \bibinfo{person}{Adrian Vladu}.}
  \bibinfo{year}{2017}\natexlab{}.
\newblock \showarticletitle{Towards deep learning models resistant to
  adversarial attacks}.
\newblock \bibinfo{journal}{\emph{arXiv preprint arXiv:1706.06083}}
  (\bibinfo{year}{2017}).
\newblock


\bibitem[\protect\citeauthoryear{Moosavi-Dezfooli, Fawzi, and
  Frossard}{Moosavi-Dezfooli et~al\mbox{.}}{2016}]%
        {moosavi2016deepfool}
\bibfield{author}{\bibinfo{person}{Seyed-Mohsen Moosavi-Dezfooli},
  \bibinfo{person}{Alhussein Fawzi}, {and} \bibinfo{person}{Pascal Frossard}.}
  \bibinfo{year}{2016}\natexlab{}.
\newblock \showarticletitle{Deepfool: a simple and accurate method to fool deep
  neural networks}. In \bibinfo{booktitle}{\emph{Proceedings of the IEEE
  conference on computer vision and pattern recognition}}.
  \bibinfo{pages}{2574--2582}.
\newblock


\bibitem[\protect\citeauthoryear{Papernot, McDaniel, Goodfellow, Jha, Celik,
  and Swami}{Papernot et~al\mbox{.}}{2017}]%
        {papernot2017practical}
\bibfield{author}{\bibinfo{person}{Nicolas Papernot}, \bibinfo{person}{Patrick
  McDaniel}, \bibinfo{person}{Ian Goodfellow}, \bibinfo{person}{Somesh Jha},
  \bibinfo{person}{Z~Berkay Celik}, {and} \bibinfo{person}{Ananthram Swami}.}
  \bibinfo{year}{2017}\natexlab{}.
\newblock \showarticletitle{Practical black-box attacks against machine
  learning}. In \bibinfo{booktitle}{\emph{Proceedings of the 2017 ACM on Asia
  conference on computer and communications security}}.
  \bibinfo{pages}{506--519}.
\newblock


\bibitem[\protect\citeauthoryear{Paszke, Gross, Chintala, Chanan, Yang, DeVito,
  Lin, Desmaison, Antiga, and Lerer}{Paszke et~al\mbox{.}}{2017}]%
        {paszke2017automatic}
\bibfield{author}{\bibinfo{person}{Adam Paszke}, \bibinfo{person}{Sam Gross},
  \bibinfo{person}{Soumith Chintala}, \bibinfo{person}{Gregory Chanan},
  \bibinfo{person}{Edward Yang}, \bibinfo{person}{Zachary DeVito},
  \bibinfo{person}{Zeming Lin}, \bibinfo{person}{Alban Desmaison},
  \bibinfo{person}{Luca Antiga}, {and} \bibinfo{person}{Adam Lerer}.}
  \bibinfo{year}{2017}\natexlab{}.
\newblock \showarticletitle{Automatic differentiation in pytorch}.
\newblock  (\bibinfo{year}{2017}).
\newblock


\bibitem[\protect\citeauthoryear{Raghunathan, Steinhardt, and
  Liang}{Raghunathan et~al\mbox{.}}{2018}]%
        {raghunathan2018certified}
\bibfield{author}{\bibinfo{person}{Aditi Raghunathan}, \bibinfo{person}{Jacob
  Steinhardt}, {and} \bibinfo{person}{Percy Liang}.}
  \bibinfo{year}{2018}\natexlab{}.
\newblock \showarticletitle{Certified defenses against adversarial examples}.
\newblock \bibinfo{journal}{\emph{arXiv preprint arXiv:1801.09344}}
  (\bibinfo{year}{2018}).
\newblock


\bibitem[\protect\citeauthoryear{Rechenberg}{Rechenberg}{1978}]%
        {rechenberg1978evolutionsstrategien}
\bibfield{author}{\bibinfo{person}{Ingo Rechenberg}.}
  \bibinfo{year}{1978}\natexlab{}.
\newblock \showarticletitle{Evolutionsstrategien}.
\newblock In \bibinfo{booktitle}{\emph{Simulationsmethoden in der Medizin und
  Biologie}}. \bibinfo{publisher}{Springer}, \bibinfo{pages}{83--114}.
\newblock


\bibitem[\protect\citeauthoryear{Ross and Doshi-Velez}{Ross and
  Doshi-Velez}{2018}]%
        {ross2018improving}
\bibfield{author}{\bibinfo{person}{Andrew~Slavin Ross} {and}
  \bibinfo{person}{Finale Doshi-Velez}.} \bibinfo{year}{2018}\natexlab{}.
\newblock \showarticletitle{Improving the adversarial robustness and
  interpretability of deep neural networks by regularizing their input
  gradients}. In \bibinfo{booktitle}{\emph{Thirty-second AAAI conference on
  artificial intelligence}}.
\newblock


\bibitem[\protect\citeauthoryear{Sen, Ravindran, and Raghunathan}{Sen
  et~al\mbox{.}}{2020}]%
        {sen2020empir}
\bibfield{author}{\bibinfo{person}{Sanchari Sen}, \bibinfo{person}{Balaraman
  Ravindran}, {and} \bibinfo{person}{Anand Raghunathan}.}
  \bibinfo{year}{2020}\natexlab{}.
\newblock \showarticletitle{Empir: Ensembles of mixed precision deep networks
  for increased robustness against adversarial attacks}.
\newblock \bibinfo{journal}{\emph{arXiv preprint arXiv:2004.10162}}
  (\bibinfo{year}{2020}).
\newblock


\bibitem[\protect\citeauthoryear{Shafahi, Najibi, Ghiasi, Xu, Dickerson,
  Studer, Davis, Taylor, and Goldstein}{Shafahi et~al\mbox{.}}{2019}]%
        {shafahi2019adversarial}
\bibfield{author}{\bibinfo{person}{Ali Shafahi}, \bibinfo{person}{Mahyar
  Najibi}, \bibinfo{person}{Mohammad~Amin Ghiasi}, \bibinfo{person}{Zheng Xu},
  \bibinfo{person}{John Dickerson}, \bibinfo{person}{Christoph Studer},
  \bibinfo{person}{Larry~S Davis}, \bibinfo{person}{Gavin Taylor}, {and}
  \bibinfo{person}{Tom Goldstein}.} \bibinfo{year}{2019}\natexlab{}.
\newblock \showarticletitle{Adversarial training for free!}. In
  \bibinfo{booktitle}{\emph{Advances in Neural Information Processing
  Systems}}. \bibinfo{pages}{3353--3364}.
\newblock


\bibitem[\protect\citeauthoryear{Sharif, Bhagavatula, Bauer, and Reiter}{Sharif
  et~al\mbox{.}}{2016}]%
        {sharif2016accessorize}
\bibfield{author}{\bibinfo{person}{Mahmood Sharif}, \bibinfo{person}{Sruti
  Bhagavatula}, \bibinfo{person}{Lujo Bauer}, {and} \bibinfo{person}{Michael~K
  Reiter}.} \bibinfo{year}{2016}\natexlab{}.
\newblock \showarticletitle{Accessorize to a crime: Real and stealthy attacks
  on state-of-the-art face recognition}. In
  \bibinfo{booktitle}{\emph{Proceedings of the 2016 acm sigsac conference on
  computer and communications security}}. \bibinfo{pages}{1528--1540}.
\newblock


\bibitem[\protect\citeauthoryear{Sharif, Bhagavatula, Bauer, and Reiter}{Sharif
  et~al\mbox{.}}{2017}]%
        {sharif2017adversarial}
\bibfield{author}{\bibinfo{person}{Mahmood Sharif}, \bibinfo{person}{Sruti
  Bhagavatula}, \bibinfo{person}{Lujo Bauer}, {and} \bibinfo{person}{Michael~K
  Reiter}.} \bibinfo{year}{2017}\natexlab{}.
\newblock \showarticletitle{Adversarial generative nets: Neural network attacks
  on state-of-the-art face recognition}.
\newblock \bibinfo{journal}{\emph{arXiv preprint arXiv:1801.00349}}
  (\bibinfo{year}{2017}), \bibinfo{pages}{1556--6013}.
\newblock


\bibitem[\protect\citeauthoryear{Simonyan and Zisserman}{Simonyan and
  Zisserman}{2014}]%
        {simonyan2014very}
\bibfield{author}{\bibinfo{person}{Karen Simonyan} {and}
  \bibinfo{person}{Andrew Zisserman}.} \bibinfo{year}{2014}\natexlab{}.
\newblock \showarticletitle{Very deep convolutional networks for large-scale
  image recognition}.
\newblock \bibinfo{journal}{\emph{arXiv preprint arXiv:1409.1556}}
  (\bibinfo{year}{2014}).
\newblock


\bibitem[\protect\citeauthoryear{Song, Shu, Kushman, and Ermon}{Song
  et~al\mbox{.}}{2018}]%
        {song2018constructing}
\bibfield{author}{\bibinfo{person}{Yang Song}, \bibinfo{person}{Rui Shu},
  \bibinfo{person}{Nate Kushman}, {and} \bibinfo{person}{Stefano Ermon}.}
  \bibinfo{year}{2018}\natexlab{}.
\newblock \showarticletitle{Constructing unrestricted adversarial examples with
  generative models}. In \bibinfo{booktitle}{\emph{Advances in Neural
  Information Processing Systems}}. \bibinfo{pages}{8312--8323}.
\newblock


\bibitem[\protect\citeauthoryear{Su, Vargas, and Sakurai}{Su
  et~al\mbox{.}}{2019}]%
        {su2019one}
\bibfield{author}{\bibinfo{person}{Jiawei Su},
  \bibinfo{person}{Danilo~Vasconcellos Vargas}, {and} \bibinfo{person}{Kouichi
  Sakurai}.} \bibinfo{year}{2019}\natexlab{}.
\newblock \showarticletitle{One pixel attack for fooling deep neural networks}.
\newblock \bibinfo{journal}{\emph{IEEE Transactions on Evolutionary
  Computation}} \bibinfo{volume}{23}, \bibinfo{number}{5}
  (\bibinfo{year}{2019}), \bibinfo{pages}{828--841}.
\newblock


\bibitem[\protect\citeauthoryear{Szegedy, Vanhoucke, Ioffe, Shlens, and
  Wojna}{Szegedy et~al\mbox{.}}{2016}]%
        {szegedy2016rethinking}
\bibfield{author}{\bibinfo{person}{Christian Szegedy}, \bibinfo{person}{Vincent
  Vanhoucke}, \bibinfo{person}{Sergey Ioffe}, \bibinfo{person}{Jon Shlens},
  {and} \bibinfo{person}{Zbigniew Wojna}.} \bibinfo{year}{2016}\natexlab{}.
\newblock \showarticletitle{Rethinking the inception architecture for computer
  vision}. In \bibinfo{booktitle}{\emph{Proceedings of the IEEE conference on
  computer vision and pattern recognition}}. \bibinfo{pages}{2818--2826}.
\newblock


\bibitem[\protect\citeauthoryear{Szegedy, Zaremba, Sutskever, Bruna, Erhan,
  Goodfellow, and Fergus}{Szegedy et~al\mbox{.}}{2013}]%
        {szegedy2013intriguing}
\bibfield{author}{\bibinfo{person}{Christian Szegedy},
  \bibinfo{person}{Wojciech Zaremba}, \bibinfo{person}{Ilya Sutskever},
  \bibinfo{person}{Joan Bruna}, \bibinfo{person}{Dumitru Erhan},
  \bibinfo{person}{Ian Goodfellow}, {and} \bibinfo{person}{Rob Fergus}.}
  \bibinfo{year}{2013}\natexlab{}.
\newblock \showarticletitle{Intriguing properties of neural networks}.
\newblock \bibinfo{journal}{\emph{arXiv preprint arXiv:1312.6199}}
  (\bibinfo{year}{2013}).
\newblock


\bibitem[\protect\citeauthoryear{Tewari and Bartlett}{Tewari and
  Bartlett}{2007}]%
        {tewari2007consistency}
\bibfield{author}{\bibinfo{person}{Ambuj Tewari} {and} \bibinfo{person}{Peter~L
  Bartlett}.} \bibinfo{year}{2007}\natexlab{}.
\newblock \showarticletitle{On the consistency of multiclass classification
  methods}.
\newblock \bibinfo{journal}{\emph{Journal of Machine Learning Research}}
  \bibinfo{volume}{8}, \bibinfo{number}{May} (\bibinfo{year}{2007}),
  \bibinfo{pages}{1007--1025}.
\newblock


\bibitem[\protect\citeauthoryear{Tram{\`e}r and Boneh}{Tram{\`e}r and
  Boneh}{2019}]%
        {tramer2019adversarial}
\bibfield{author}{\bibinfo{person}{Florian Tram{\`e}r} {and}
  \bibinfo{person}{Dan Boneh}.} \bibinfo{year}{2019}\natexlab{}.
\newblock \showarticletitle{Adversarial training and robustness for multiple
  perturbations}. In \bibinfo{booktitle}{\emph{Advances in Neural Information
  Processing Systems}}. \bibinfo{pages}{5858--5868}.
\newblock


\bibitem[\protect\citeauthoryear{Tramer, Carlini, Brendel, and Madry}{Tramer
  et~al\mbox{.}}{2020}]%
        {tramer2020adaptive}
\bibfield{author}{\bibinfo{person}{Florian Tramer}, \bibinfo{person}{Nicholas
  Carlini}, \bibinfo{person}{Wieland Brendel}, {and}
  \bibinfo{person}{Aleksander Madry}.} \bibinfo{year}{2020}\natexlab{}.
\newblock \showarticletitle{On adaptive attacks to adversarial example
  defenses}.
\newblock \bibinfo{journal}{\emph{arXiv preprint arXiv:2002.08347}}
  (\bibinfo{year}{2020}).
\newblock


\bibitem[\protect\citeauthoryear{Tram{\`e}r, Kurakin, Papernot, Goodfellow,
  Boneh, and McDaniel}{Tram{\`e}r et~al\mbox{.}}{2017}]%
        {tramer2017ensemble}
\bibfield{author}{\bibinfo{person}{Florian Tram{\`e}r}, \bibinfo{person}{Alexey
  Kurakin}, \bibinfo{person}{Nicolas Papernot}, \bibinfo{person}{Ian
  Goodfellow}, \bibinfo{person}{Dan Boneh}, {and} \bibinfo{person}{Patrick
  McDaniel}.} \bibinfo{year}{2017}\natexlab{}.
\newblock \showarticletitle{Ensemble adversarial training: Attacks and
  defenses}.
\newblock \bibinfo{journal}{\emph{arXiv preprint arXiv:1705.07204}}
  (\bibinfo{year}{2017}).
\newblock


\bibitem[\protect\citeauthoryear{Uesato, O'Donoghue, Oord, and Kohli}{Uesato
  et~al\mbox{.}}{2018}]%
        {uesato2018adversarial}
\bibfield{author}{\bibinfo{person}{Jonathan Uesato}, \bibinfo{person}{Brendan
  O'Donoghue}, \bibinfo{person}{Aaron van~den Oord}, {and}
  \bibinfo{person}{Pushmeet Kohli}.} \bibinfo{year}{2018}\natexlab{}.
\newblock \showarticletitle{Adversarial risk and the dangers of evaluating
  against weak attacks}.
\newblock \bibinfo{journal}{\emph{arXiv preprint arXiv:1802.05666}}
  (\bibinfo{year}{2018}).
\newblock


\bibitem[\protect\citeauthoryear{Wang and Zhang}{Wang and Zhang}{2019}]%
        {wang2019bilateral}
\bibfield{author}{\bibinfo{person}{Jianyu Wang} {and} \bibinfo{person}{Haichao
  Zhang}.} \bibinfo{year}{2019}\natexlab{}.
\newblock \showarticletitle{Bilateral adversarial training: Towards fast
  training of more robust models against adversarial attacks}. In
  \bibinfo{booktitle}{\emph{Proceedings of the IEEE International Conference on
  Computer Vision}}. \bibinfo{pages}{6629--6638}.
\newblock


\bibitem[\protect\citeauthoryear{Wang, Zou, Yi, Bailey, Ma, and Gu}{Wang
  et~al\mbox{.}}{2019}]%
        {wang2019improving}
\bibfield{author}{\bibinfo{person}{Yisen Wang}, \bibinfo{person}{Difan Zou},
  \bibinfo{person}{Jinfeng Yi}, \bibinfo{person}{James Bailey},
  \bibinfo{person}{Xingjun Ma}, {and} \bibinfo{person}{Quanquan Gu}.}
  \bibinfo{year}{2019}\natexlab{}.
\newblock \showarticletitle{Improving adversarial robustness requires
  revisiting misclassified examples}. In
  \bibinfo{booktitle}{\emph{International Conference on Learning
  Representations}}.
\newblock


\bibitem[\protect\citeauthoryear{Xiao, Li, Zhu, He, Liu, and Song}{Xiao
  et~al\mbox{.}}{2018a}]%
        {xiao2018generating}
\bibfield{author}{\bibinfo{person}{Chaowei Xiao}, \bibinfo{person}{Bo Li},
  \bibinfo{person}{Jun-Yan Zhu}, \bibinfo{person}{Warren He},
  \bibinfo{person}{Mingyan Liu}, {and} \bibinfo{person}{Dawn Song}.}
  \bibinfo{year}{2018}\natexlab{a}.
\newblock \showarticletitle{Generating adversarial examples with adversarial
  networks}.
\newblock \bibinfo{journal}{\emph{arXiv preprint arXiv:1801.02610}}
  (\bibinfo{year}{2018}).
\newblock


\bibitem[\protect\citeauthoryear{Xiao, Zhu, Li, He, Liu, and Song}{Xiao
  et~al\mbox{.}}{2018b}]%
        {xiao2018spatially}
\bibfield{author}{\bibinfo{person}{Chaowei Xiao}, \bibinfo{person}{Jun-Yan
  Zhu}, \bibinfo{person}{Bo Li}, \bibinfo{person}{Warren He},
  \bibinfo{person}{Mingyan Liu}, {and} \bibinfo{person}{Dawn Song}.}
  \bibinfo{year}{2018}\natexlab{b}.
\newblock \showarticletitle{Spatially transformed adversarial examples}.
\newblock \bibinfo{journal}{\emph{arXiv preprint arXiv:1801.02612}}
  (\bibinfo{year}{2018}).
\newblock


\bibitem[\protect\citeauthoryear{Xie, Wang, Zhang, Ren, and Yuille}{Xie
  et~al\mbox{.}}{2017}]%
        {xie2017mitigating}
\bibfield{author}{\bibinfo{person}{Cihang Xie}, \bibinfo{person}{Jianyu Wang},
  \bibinfo{person}{Zhishuai Zhang}, \bibinfo{person}{Zhou Ren}, {and}
  \bibinfo{person}{Alan Yuille}.} \bibinfo{year}{2017}\natexlab{}.
\newblock \showarticletitle{Mitigating adversarial effects through
  randomization}.
\newblock \bibinfo{journal}{\emph{arXiv preprint arXiv:1711.01991}}
  (\bibinfo{year}{2017}).
\newblock


\bibitem[\protect\citeauthoryear{Xu, Evans, and Qi}{Xu et~al\mbox{.}}{2017}]%
        {xu2017feature}
\bibfield{author}{\bibinfo{person}{Weilin Xu}, \bibinfo{person}{David Evans},
  {and} \bibinfo{person}{Yanjun Qi}.} \bibinfo{year}{2017}\natexlab{}.
\newblock \showarticletitle{Feature squeezing: Detecting adversarial examples
  in deep neural networks}.
\newblock \bibinfo{journal}{\emph{arXiv preprint arXiv:1704.01155}}
  (\bibinfo{year}{2017}).
\newblock


\bibitem[\protect\citeauthoryear{Zhang, Zhang, Lu, Zhu, and Dong}{Zhang
  et~al\mbox{.}}{2019b}]%
        {zhang2019you}
\bibfield{author}{\bibinfo{person}{Dinghuai Zhang}, \bibinfo{person}{Tianyuan
  Zhang}, \bibinfo{person}{Yiping Lu}, \bibinfo{person}{Zhanxing Zhu}, {and}
  \bibinfo{person}{Bin Dong}.} \bibinfo{year}{2019}\natexlab{b}.
\newblock \showarticletitle{You only propagate once: Painless adversarial
  training using maximal principle}.
\newblock \bibinfo{journal}{\emph{arXiv preprint arXiv:1905.00877}}
  \bibinfo{volume}{2} (\bibinfo{year}{2019}).
\newblock


\bibitem[\protect\citeauthoryear{Zhang, Yu, Jiao, Xing, Ghaoui, and
  Jordan}{Zhang et~al\mbox{.}}{2019a}]%
        {zhang2019theoretically}
\bibfield{author}{\bibinfo{person}{Hongyang Zhang}, \bibinfo{person}{Yaodong
  Yu}, \bibinfo{person}{Jiantao Jiao}, \bibinfo{person}{Eric~P Xing},
  \bibinfo{person}{Laurent~El Ghaoui}, {and} \bibinfo{person}{Michael~I
  Jordan}.} \bibinfo{year}{2019}\natexlab{a}.
\newblock \showarticletitle{Theoretically principled trade-off between
  robustness and accuracy}.
\newblock \bibinfo{journal}{\emph{arXiv preprint arXiv:1901.08573}}
  (\bibinfo{year}{2019}).
\newblock


\end{thebibliography}

\end{document}